\documentclass[conference]{IEEEtran}
\IEEEoverridecommandlockouts

\usepackage{cite}
\usepackage{amsmath,amssymb,amsfonts}
\usepackage{soul}
\usepackage{graphicx}
\usepackage{algorithmic}
\usepackage{textcomp}
\usepackage{xcolor}

\usepackage{color}
\usepackage{subfigure}
\usepackage[T1]{fontenc}

\usepackage[bookmarks=false]{hyperref}
\usepackage{caption}
\usepackage{subcaption}
\usepackage{wrapfig}
\usepackage{tabularx}

\usepackage{adjustbox}
\usepackage{multirow}
\usepackage{listings}
\usepackage{tabularx}
\usepackage{booktabs}
\usepackage{array}
\usepackage[symbol]{footmisc}
\def\BibTeX{{\rm B\kern-.05em{\sc i\kern-.025em b}\kern-.08em
    T\kern-.1667em\lower.7ex\hbox{E}\kern-.125emX}}

\newcommand{\changed}[1]{\textcolor{black}{{#1}}}
\newcommand{\added}[1]{\textcolor{black}{{#1}}}
\newcommand{\adding}[1]{\textcolor{black}{{#1}}}

\begin{document}

\title{Online Authentication Habits of Indian Users}

\author{\IEEEauthorblockN{Pratyush Choudhary, Subhrajit Das, Mukul Paras Potta, Prasuj Das, and Abhishek Bichhawat}
\IEEEauthorblockA{\textit{Indian Institute of Technology Gandhinagar} \\
Gandhinagar, India \\
\{pratyush.choudhary, subhrajit.das, mukul.potta, dasprasuj, abhishek.b\}@iitgn.ac.in}
}

\maketitle

\begin{abstract}
Passwords have been long used as the primary authentication method for web services. Weak passwords used by the users have prompted the use of password management tools and two-factor authentication to ensure better account security. While prior studies have studied their adoption individually, none of these studies focuses particularly on the Indian setting, which is culturally and economically different from the countries in which these studies have been done in the past. To this end, we conducted a survey with 90 participants residing in India to better understand the mindset of people on using password managers and two-factor authentication (2FA). 

Our findings suggest that a majority of the participants have used 2FA and password managers in some form, although they are sometimes unaware of their formal names. While many participants used some form of 2FA across all their accounts, browser-integrated and device-default password managers are predominantly utilized for less sensitive platforms such as e-commerce and social media rather than for more critical accounts like banking. The primary motivation for using password managers is the convenience of auto-filling. However, some participants avoid using password managers due to a lack of trust in these tools. Notably, dedicated third-party applications show low adoption for both password manager and 2FA.

Despite acknowledging the importance of secure password practices, many participants still reuse passwords across multiple accounts, prefer shorter passwords, and use commonly predictable password patterns. Overall, the study suggests that Indians are more inclined to choose default settings, underscoring the need for tailored strategies to improve user awareness and strengthen password security practices.
 
\end{abstract}


\section{Introduction}
\label{introduction}
Passwords are used as a standard authentication method for services. People are actively managing 16 to 26 protected online accounts \cite{pearman2019people, stobert2018password, Alodhyani2020_trust_transparency}, and the average workplace password manager user has 191 accounts in general~\cite{lastpass_blog_workplace}. Managing these many protected online accounts has led to unsafe practices of password reuse or selection of weak passwords \cite{gaw2006password, pearman2017let}. In conjunction with frequent data breaches~\cite{aadhaar_tribune,yahoo_reuters,dubsmash_mozilla}, password reuse or weak passwords increase the risk of users losing their accounts to an adversary~\cite{pearman2017let, pearman2019people,florencio2007large}. 

Password managers were introduced to mitigate the issue of password reuse by providing a way to remember "strong" passwords. Some of these password managers offer additional features like password generation to help generate strong passwords, adding randomness without using any of the user's personal information, making them harder to be guessed~\cite{pii_passwords}.

An alternative approach to ensuring the security of accounts, even when users use weak passwords, is using two-factor authentication (2FA). 2FA relies on an additional tool or technique beyond the primary authentication method of just entering the password to log into any account. For example, a password followed by a one-time pin (OTP) sent to your phone provides an additional layer of security, i.e., even if the password is known to the adversary, they cannot gain access to the account without knowing the OTP. 

While password managers and 2FA offer enhanced security, their usability remains challenged~\cite{pearman2019people, gaw2006password, pearman2017let}. 
Prior works~\cite{whitty_password_sharing,tam2010psychology} \changed{have shown that users often understand good and bad password practices but still engage in poor password management due to the lack of immediate negative effects. Additionally, despite scoring high on self-monitoring regarding cybersecurity practices, younger individuals are more likely to share passwords.}

With the internet becoming easily accessible in India~\cite{kemp_2024}, more users are creating online accounts to connect to the web. However, the awareness and adoption of good practices with respect to passwords amongst the Indian user base remains unclear. 

\added{Although there have been some studies on cybersecurity/password behaviour and password managers in culturally similar contexts to India like Sri Lanka~\cite{fernandoart,fernandoexploring}, Bangladesh~\cite{khan2018security,haque2013password} and Pakistan~\cite{al2020women}, none of them have been performed in the Indian context. Amongst the prior studies, some of them studied password managers briefly, but none of them explored 2FA.} 

\paragraph{Goal of this work}
Our focus in this work is to investigate the perceptions of Indian users of different authentication mechanisms and the convenience of different security and privacy tools, mainly focused on password managers and 2FA. To this end, we conducted a survey with 90 Indian participants from diverse occupational, cultural and educational backgrounds. Through a structured survey, the study aims to investigate Indian users' awareness of services that assist in easing the process of remembering strong and complex passwords through password managers and 2FA techniques. Further, we would like to understand the perception of Indian users when using these tools,  why they do (not) prefer to use them for their login processes, and how their perception changes across different scenarios.

In particular, we investigated users' password habits and whether they have been exposed to password managers and 2FA in their various forms. We also investigated their views on password manager features and security, different 2FA methods and their preferences for them. Furthermore, we explored whether their mindset changes when they are accessing different types of websites like banking, e-commerce or social media.

Our findings suggest an awareness of security amongst the participants; however, their habits do not align with their security beliefs. The participants reported having used password managers and 2FA in some or the other form in their day-to-day lives, but mostly if it was set as the default option. The majority used 2FA in the form of OTPs (as mandated by the web services) and used browser built-in password-saving features. Convenience was a major reason for the preference for using password managers, while the users found the use of 2FA cumbersome. Perceived comfort or ease of use was high when password managers and 2FA were used after a tutorial, indicating that training people on using these tools may increase their chances of adopting them. Interestingly though, the majority of the participants did not trust password managers with credentials for high-sensitivity applications but felt safer when using 2FA for their online activities. The participants also did not prefer using services from vendors involved in a data breach in the past, indicating that their trust was directly affected by the security practices of the organization whose services were used.

The subsequent sections of the paper are organized as follows: 

Section~\ref{relatedwork} reviews related works. The overall survey design methodology is described in Section~\ref{methodology}, and the observations and results are presented in Section~\ref{results}. Section~\ref{discussion} offers a discussion of the findings, and Section~\ref{conclusion} concludes. The survey instrument and collected anonymous data are available at \url{https://osf.io/9e3yu/?view_only=54e48f58e9774031870f62e230a21034}.
\section{Related Works} 
\label{relatedwork}

In recent years, extensive research has highlighted numerous usability challenges associated with password managers and 2FA~\cite{fagan2017investigation, chiasson2006usability, Alodhyani2020_trust_transparency, ion2015no, reese2019usability, acemyan_google, das_2018_fido}, but none of them study the usability and perception of both password managers and 2FA together. While password managers help in creating and remembering complex secure passwords, 2FA provides the flexibility of having a weaker password (although not recommended) due to the second-factor authentication. 
Below, we discuss prior works related to our study, categorizing them into three themes: passwords, password managers, and two-factor authentication.

\subsubsection{Passwords}

While studies show that many implementations of passwords are not without usability issues like varying length requirements, poor storage (both client-side and server-side) and memorability ~\cite{adams-1999, jobusch-1989, morris-1979, proctor-2002, yan-2004, wash_frequent_pass}, they are yet to be replaced by newer methods~\cite{bonneau2012} due to their widespread usage~\cite{herley-2009}. Though they are flawed, some studies claim that passwords -- given that they are implemented properly -- are the best suited to the use cases in the wild, largely attributing it to the ease of deployment~\cite{bonneau2010password, herley2012}. Research on passwords has continued, and they have been subject to multiple studies where they were proven to be vulnerable to various attacks. But due to the absence of similarly usable alternatives, an arms race has been going on between security experts trying to keep improving the strength of passwords, suggesting better policies and practices, and the bad actors with ever-increasing computation power and better attack techniques at their disposal~\cite{herley-2009, herley2012, kuo-2006, malone-2012}.

There have been studies aiming to evaluate the strength of a password quantitatively. Some of those studies suggested the following: using naive entropy (the higher, the better; but doesn't take password randomness into account), popularity (which requires tracking the passwords used, a threat to security), number of guesses (which differ across the methods used from guessing the password)~\cite{kelley-2011, schechter-2010, ur-2015}.

\subsubsection{Password managers}
Password managers enable users to manage a large number of accounts and have randomised passwords with quality-of-life features like password auto-fill and cross-device sync. Various studies have explored password manager usage trends, the motivations of users, and the common pitfalls of using or not using password managers, the majority of which were from the US or Canada.

Ion et al.~\cite{ion2015no} found out 
that only 24\% of "non-security experts" used password managers while others wrote down passwords, remembered, or reused them. Many expressed distrust in password managers and poor usability as reasons for their low adoption. 

\adding{The lack of trust in password managers and poor usability were also confirmed by other works}~\cite{stobert2018password, fagan2016they, fagan2017investigation, chiasson2006usability, Alodhyani2020_trust_transparency}. 

The people who use password managers did so due to the added convenience of autofill, with another reason being added security~\cite{fagan2016they, fagan2017investigation}; however, they did not completely understand their security benefits as many reported they did not trust a password manager for sensitive accounts~\cite{fagan2017investigation}.

Recent studies~\cite{stobert2016expert,stobert2018password,pearman2019people} investigated the choice of using a browser/device built-in managers vs. standalone applications and reported that built-in password managers are used mainly for convenience and standalone ones for security reasons.

\subsubsection{Two-factor authentication}
As there are various means to 2FA, most studies have either focused on a specific method or have compared them for usability and user perception. 

Cristofaro et al.~\cite{CristofaroDFN13} studied the usability and perceived usability of three popular 2FA methods: security token codes, email/SMS OTPs, and dedicated third-party authentication apps (e.g., Google Authenticator). Their findings suggest that 2FA is generally perceived as usable, with ease of use, cognitive effort, and trustworthiness being key factors affecting usability. The study also revealed that individual characteristics, such as age and gender, influence the perception of 2FA more than the technology itself. Several  works~\cite{das_2018_fido,acemyan_google,weidman_byod_employee,reese2019usability} studied the use of hardware tokens and other methods for 2FA and reported that users found them difficult to use and preferred the default devices and settings for the same.

In 2022, Marky et al.~\cite{marky2022nah} conducted a study through semi-structured interviews with 42 participants from European countries (Germany, France, and the UK) and from Asian countries (China, Japan, Taiwan, and South Korea) on user perceptions and experience of 2FA. The study confirmed previous work and revealed that users often find 2FA setup procedures too long and complicated, affecting their integration into daily routines. Users also faced challenges of poor integration across different systems and the inconvenience of carrying multiple 2FA devices. 
However, there has been no significant study conducted on Indian users in the past on 2FA, although 2FA has been there for a while either due to mandatory compliance (e.g. banking~\cite{rbi2025vision}) or recommendation by the web services for extra security. In this work, we investigate the perception of 2FA amongst Indian users.
\section{Methodology} 
\label{methodology}

To investigate the awareness, perception, and convenience of password managers and 2FA of Indian users, we conducted a structured survey; we obtained ethical clearance from our Institutional Ethics Committee for the same. 

\subsection{Participant Recruitment}

\added{Participants residing in India were recruited through a comprehensive strategy utilizing multiple channels to diversify and recruit as many users as possible. Recruitment advertisements were circulated via various social media platforms, including WhatsApp, LinkedIn, and email. Due to low participation through these platforms, we leveraged our contacts, friends, and some university groups to maximize participation.}  

\added{The survey took place from the end of January to the beginning of March 2024, over approximately six weeks, which was the timeline for data collection as per the approval received from the ethics committee.} 

Each participant was asked to fill out a consent form at the beginning of the survey. The questionnaire included attention-check questions to ensure participants' engagement when responding. \added{On average, the participants took $\sim$30 minutes to fill out the survey.} Of a total of 121 responses, 31 invalid responses were excluded from the dataset as they failed the attention-check questions, resulting in 90 valid responses.  The participants included a mix of people with the main criteria for recruitment being that they must be 18 or older, and they must have created or used at least one password in the past. To obtain a diverse dataset, we did not require them to know about password managers or 2FA. \changed{Some participants had difficulty understanding certain English terms and reached out to us for clarification. We responded by explaining these terms using simpler language, occasionally in their local language, or by providing context through explanatory examples.}  We did not collect any personal or identifying information other than participants' payment details (UPI ID) for compensation, which was entirely optional and provided at Rs. 100 per hour. Once the compensation was processed, the payment information was removed from our database.

\subsection{Pilot Survey}
We conducted a pilot study involving five individuals and incorporated their feedback into refining the questionnaire for our actual study. An important piece of feedback received as part of the pilot study was that the participants wanted to refrain from responding to certain questions because "\textit{... this information is about the passwords themselves ...}." \changed{Additionally, our institute's review committee required that participation be entirely voluntary, allowing participants to skip any questions. To accommodate this policy and to increase participation, we included a "Prefer not to say" option for multiple questions throughout the questionnaire. While we did not store any personally identifiable information, this option helped ensure that participants felt comfortable with their responses.}
Another concern from the participants was that there was no option to convey uncertainty as to whether they used these mechanisms since they asked questions like, "What would be classified as two-factor authentication?" and "Is OTP one of them?" among others. To alleviate this, we added the option "Maybe" to the responses, which indicated that the participants thought they were using a particular service but did not know whether it qualified as either a manager or a 2FA technique.

\subsection{Study Design}
The survey questionnaire had four major components. The first component focused on questions related to the participant's password habits and estimating their digital footprint, e.g., the number of online accounts owned, types of devices used for web access, the number of accounts created by others, frequency of password logins, the number of different passwords used daily/monthly, instances of shared passwords across accounts, experiences of forgotten passwords, methods of recording passwords, frequency of password changes, common patterns in password creation, preferred password length, and the use of fixed templates for creating passwords. The second component dealt with the awareness, usage and perceptions of participants towards password managers, while the third component dealt with 2FA schemes. The final component collected demographic information and contained a survey rating and feedback section.

The components dealing with password managers and 2FA had the following flow: 
Based on whether the participants had used either of these techniques, we presented different sets of questions. If the participants responded positively to using either password managers or 2FA or both, we surveyed their experience using them and their perceptions. Irrespective of the usage of these mechanisms by the participants in the past, we required them to watch informative videos \cite{youtubeWhatTwoFactor,youtubePasswordManagers,youtubeEnableFactor,youtubeBitwardensPassword} about that technique, specifically, how the technique works and how to set it up for their accounts. After the participants had watched the videos, we enquired about their perceptions of safety, convenience, and security.

The survey instrument consisted of a total of 70 questions: six for demographics, 14 for digital footprint, 18 for 2FA, 25 for password managers, and seven for both 2FA and password managers. Due to the absence of Prolific or a similar survey platform in India, we presented the questionnaire using simple forms distributed to participants over email. Additionally, a printed version of the questionnaire, augmented with QR codes linking to the informative videos, was disseminated to facilitate participation among a broader audience.

\subsection{Analysis}

From the data we collected and cleaned, we computed the following statistics and tests, which will be mentioned further in the paper where relevant: 

\begin{itemize}
   
    \item Cramer's V with bias correction: a symmetric measure of association between two categorical variables with two or more categories. We implemented the bias correction from ~\cite{bergsma-2013} to compute the statistic. We use this to identify correlations among answers to questions regarding perceptions and demographics.
    
    \item \added{Kruskal-Wallis H test ~\cite{kruskal-1952}: This tests the null hypothesis that the distributions from which the samples are taken are the same. This is used when comparing three or more independent samples to understand their underlying distributions. Since this test can only indicate whether there is some pair of samples whose distributions differ, we add Dunn's test ~\cite{dunn-1964} with Holm's Correction ~\cite{29def780-e117-38f0-8afb-edf384af3fad} as a post hoc to identify the pairs that are significantly different.} This was used to report the users' preference rating for various 2FA methods in Sec. IV-C.
    \item \added{Wilcoxon's signed-rank test ~\cite{wilcoxon-1945}: This tests the null hypothesis that the paired samples are taken from the same distribution and used in a scenario where we have two related samples.} This was used to report the users' preference rating for various 2FA methods in Sec. IV-C.
\end{itemize}

\subsection{Participant Demographics}
Out of the 90 participants, 72 identified as male, 16 as female, and two chose not to identify their gender. Around 49\% of the participants were in the age group 18-25, 32\% were 26-35 years of age, 15\% were 36-45, while 1\% and 3\% of them were 46-55 and more than 56 years of age. 
Most of our participants were well-educated: about 50\% had a post-graduate degree, and 33.3\% had an undergraduate degree. 15.7\% of the participants reported having a high school diploma or a lower education. Only three participants identified as IT professionals, while twelve participants worked in technical fields, and twelve participants occupied administrative or financial roles. The majority of our participants were students. 
Thirty-eight participants self-identified as belonging to either rural or semi-urban regions, while the remaining participants came from an urban setting (see Table \ref{tab:participants_demographics} for more demographic insights).

\begin{table}[h!]
\centering
\tiny
\renewcommand{\arraystretch}{1} 
\setlength{\tabcolsep}{1pt} 
\begin{tabular}{|p{1.1cm}|p{0.3cm}|p{0.7cm}|p{0.3cm}|p{1.3cm}|p{0.3cm}|p{1.8cm}|p{0.3cm}|p{1.0cm}|p{0.3cm}|}
\hline
\multicolumn{2}{|c|}{\textbf{Gender}} & 
\multicolumn{2}{c|}{\textbf{Age}} & 
\multicolumn{2}{c|}{\textbf{Highest Education }} &
\multicolumn{2}{c|}{\textbf{State of Residence}~\cite{ISCS_GOI,NEC_GOI}} & 
\multicolumn{2}{c|}{\textbf{Type of Region}} \\ \hline
Male & 72 & 18--25 & 44 & High Sch./Below & 14 & Southern States & 13 & Rural & 17 \\ 
Female & 16 & 26--35 & 29 & Diploma Degree & 1 & Western States & 25 & Semi-Urban & 21 \\ 
Not Disclosed & 2 & 36--45 & 13 & Bachelors Degree & 30 & North-Eastern States & 1 & Urban & 52 \\ 
& & 46--55 & 1 & Masters Degree & 33 & Eastern States & 25 & &\\ 
& & $\geq$ 56 & 3 & Doctorate Degree & 12 & Central States & 11 & &\\ 
& & & & & & Northern States & 12 & &\\
& & & & & & Not Disclosed & 3 & &\\ \hline
\end{tabular}
\caption{Participants' demographics (Total $n=90$)}
\label{tab:participants_demographics}
\vspace{-3em}
\end{table}

\subsection{Limitations}

\added{We acknowledge that our sample of 90 participants may not fully represent India's large population of over 1.4 billion ~\cite{unIndiaOvertake}. Despite our efforts, recruiting more participants proved challenging. However, this sample provides a foundational understanding of the preferences and behaviours within the Indian population.}
Almost half of the participants were from the age group 18-25, while a majority (80\%) of the participants identified themselves as male. While we tried to diversify as much as possible, we could not find female and older participants agreeing to take the survey. Most of the participants are either from the eastern or the western states in India; we received only one participant from the north-eastern states, twelve from the northern states, eleven from the central states, and 13 from the southern states. Although we have representations from the rural regions, most of our participants are fairly well-educated, which may have introduced a bias against the less educated people from the society in the survey. We found it difficult to recruit rural users for the survey even though compensation was offered.

As this was a survey concerning security, the participants' responses may have been biased, too, as they may have wanted to either come across as knowledgeable or be cautious and not reveal their actual behaviours. Also, since the survey was self-reported, participants may have under- or over-estimated their responses to certain questions.

We did not ask for the participants' proficiency in using digital devices and familiarity with security practices, in general, apart from the security mechanisms being focused on. The survey did not probe the participants on whether the usage of the mechanisms discussed in the study was voluntary or forced and how extensively they used these mechanisms. Quite a few participants in the final survey chose the "Prefer not to say" option in multiple scenarios, indicating their reluctance to share information about their authentication habits, which may have affected our findings and results.
\section{Results} 
\label{results}
The following observations were made after performing the analyses mentioned in Sec. III-D \added{(the relevant statistics are mentioned wherever a correlation is reported)} on the responses after cleaning and removing invalid responses based on attention-checking questions. Additionally, we documented the overall awareness, usage and convenience of the tools.
\subsection{Participants' Digital Footprint and Password Habits}
We start by discussing the digital presence of the participants and their habits when creating passwords on the internet. The majority of the participants (60\%) had more than six accounts, while at least half of them reported having more than ten online accounts. They normally access these accounts over a mobile phone or a PC while rarely using other devices to access the same. Sixty-five (72.2\%) participants indicated that all these accounts were created by themselves without any help from others, while the remaining participants had taken help at least once when creating these accounts online. 

Eighty-one (90\%) participants reported that they had forgotten one of their passwords at some time in the past, although 71 (78.9\%) participants reported using only 1-5 passwords on a daily basis for accessing their accounts. Forty-five (50\%) of the participants reported storing their account passwords digitally, out of which 17 (37.8\% of 45) of them stored the passwords for all of their accounts. Thirty-two (35.6\%) participants reported storing the passwords physically.

Fifty-five (61.1\%) participants reported changing their passwords at least once a year, out of which 25 participants changed their password on a quarterly basis, and seven reported doing this on a monthly basis. Sixteen (17.8\%) participants changed their passwords variably based on when they forgot their password or when mandated by the web service to be changed. Only one of them changed the passwords whenever it was possibly leaked as part of a data breach. Forty-eight (53.3\%) participants reported having a template or set of standard rules for creating new passwords.

\textbf{Password Practices:}
\changed{Most participants used common password patterns frequently found in leaked passwords ~\cite{ncscPasswords}, such as personal names (45.6\%), birthdays (41.1\%), abbreviations of personal information (26.7\%), phone numbers (26.7\%), and basic terms with numeric suffixes like "password123" (26.7\%), among others}.  

Additionally, 73 (81.1\%) participants preferred passwords that are 8-15 characters long, eleven (12.2\%) preferred passwords shorter than 8 characters, and only three (3.3\%) participants preferred passwords longer than 15 characters. Over half of them (48 participants) admitted to reusing passwords for more than $25\%$ of their accounts. 

\textbf{Importance of security practices across professions:} 
Almost all of the participants (86 participants or 95.6\%) consider it important to stay updated on security practices. This sentiment is consistent regardless of their occupations. 
\changed{However, this number may have been affected by the fact that the participants were aware that the survey was related to passwords and 2FA, which may have introduced a social desirability bias, 
as many participants reported updating their knowledge of attacks very infrequently.}

\subsection{Password managers}

\subsubsection{Use of Password Managers} 
Out of the 90 participants, 41 (45.6\%) participants reported using \textit{any} password managers in the past, of whom 35 (38.9\%) said that they were currently using password managers. Most of the participants use device-default (26 participants or 28.9\%) or browser-built-in password managers (18 participants or 20\%), while very few have used dedicated third-party password managers (five participants or 5.6\%). One participant said, "Built my own manager". This reflects a trend towards leveraging built-in security features for convenience. \added{Our analysis shows that this correlation between the usage of password managers and the perception of convenience is significant $(V = 0.804, p < 0.001)$.}
Interestingly, 87 (96.7\%) participants reported observing prompts on browsers for saving passwords, and 63 (70\%) of them chose to use this browser feature, indicating that some of the participants may not be aware that the said feature is a password manager. 
Seventy-five (83.3\%) of them later recalled encountering saved password suggestions from the browsers, suggesting that some of them were unconsciously using this browser feature.
44.4\% (40) of the participants acknowledged saving their social media passwords, while 51.1\% (46) of them acknowledged saving passwords for e-commerce websites using a password manager; however, 83.3\% (75) participants stated that they did not save their online banking passwords. 
 
\added{We selected banking, social media, and e-commerce because they represent a range of security perceptions: banking denotes high-security perception ~\cite{businessperspectivesSecurityPerception}, social media lower ~\cite{Yadav2022}, and e-commerce ~\cite{ecommereSecPercp} falls in between. This choice targets the most common daily use case scenarios ~\cite{grabonInternetUsers} and simplifies the analysis. While we did not consider healthcare, government, or educational websites, we aimed to streamline the study by focusing on these three categories. We acknowledge that including more categories could have provided a more comprehensive analysis. }

\subsubsection{Preferences and Perceptions when using Password Managers}
Of the participants who reported currently using password managers (n = 35), 29 (82.9\%) specifically used these tools because they found the auto-filling feature convenient, and 22 (62.9\%) of them cited the easy accessibility of passwords as the reason for using them.  
This indicates a significant preference for features that enhance ease of use and efficiency. Eighteen (51.4\%) participants indicated that they did not want to remember their passwords, while ten (28.6\%) said that their passwords were too complex to remember. 

The major reasons for the participants who did not currently use password managers (n = 55) were a lack of trust in password managers (26 participants or 47.3\%) and the ability to remember all their passwords (21 participants or 38.2\%).   
Thirteen (23.6\%) participants reported using easy-to-remember passwords, four (7.3\%) participants reported reusing the passwords, and nine (16.4\%) of them wrote their passwords non-digitally. Five (9.1\%) participants used password managers in the past but stopped using them, citing reasons such as they instead started writing their passwords non-digitally and remembering all their passwords as the password managers were not intuitive or were hard to use.

After the participants were shown the tutorial \cite{youtubePasswordManagers} on what password managers were, different types of managers and the benefits of using them, 50 (55.6\%) participants agreed that they would prefer using password managers in the future, while 16 (17.8\%) participants indicated that they were still not sure about using the password managers. The remaining participants (24 participants or 26.7\%) did not want to use the password managers even after the tutorial explained the benefits of using them.

\subsection{Two-factor authentication}
\subsubsection{Use of 2FA}
Most participants (85 participants or 94.4\%) were aware of or had used some form of 2FA.
 
Although some of them were initially unfamiliar with the term "two-factor authentication", when given an example (as a screenshot and some textual description) --- such as logging into an Amazon account by first entering the account password followed by an OTP sent to their phone --- many of them acknowledged they had used it.

Additionally, out of the 90 participants, 85 (94.4\%) reported having used 2FA for some online activities, with 46 (51.1\%) participants using it for social media applications and 72 (80\%) participants using it for online banking. 
 
Sixty-six (73.3\%) participants used 2FA for both social media and online banking.

\subsubsection{Different types of 2FA techniques used}
 OTPs and PINs are the most widely used factors for 2FA, with 85 (94.4\%) and 59 (65.6\%) participants reporting using them. Authentication apps (45 participants or 50\%) and biometrics (39 participants or 43.3\%) have also been used for 2FA, while hardware keys were the least used.
 Some of the participants chose the "others" option, citing they had used "recovery key and email", "SMS", and "email OTP + OTP on phone", which were already accounted for within the categories of "Hardware Key", "PINs" and "OTPs". \adding{Though OTPs via email may be argued not to be a true second factor since they're password-protected, many users in India receive them on mobile devices without re-entering their password—a method widely used by banks and web services across the country.}

\subsubsection{Preferences and Perceptions when using 2FA} 
When asked to rate their preference (on a scale from 0 to 5) for various authentication methods, participants showed the highest preference for biometrics, with an average score of 3.39 \added{($\sigma$=1.796)}.  
This was followed by OTP through SMS (3.34 \added{, $\sigma$=1.566}), OTP through email (3.18, \added{$\sigma$=1.496}), authentication apps like Google Authenticator and DuoMobile (2.90, \added{$\sigma$=1.629}), and PIN (2.89, \added{$\sigma$=1.659}). The least preferred option was hardware keys, with an average score of 2.20 \added{($\sigma$=1.762)}.

We performed the Kruskal-Wallis H test with Dunn's test as a post hoc. The Kruskal-Wallis test result indicated that there is a statistically significant difference in the distributions of the ratings ($H = 30.387, p = 0.00001$), with Dunn's test after Holm's corrections revealing that the differences for the following pairs of 2FA methods are significant: hardware keys with email OTP, SMS OTP and biometrics, respectively. 
Additionally, we performed the Wilcoxon signed-rank test on pairs of 2FA methods. With the alternative hypothesis that the distribution underlying the differences is stochastically greater than a distribution symmetric about zero, we find that the higher preference for OTPs and biometrics and the lower preference for hardware keys are statistically significant. In essence, this indicates that Indian users tend to prefer OTPs and biometrics but are skeptical about hardware keys.

In case the primary 2FA-linked device is unavailable, the most preferred alternative is receiving an OTP via the registered email, making it a trusted recovery method.

90\% of the participants (81) agree that 2FA provides security while authenticating into their account, while 75.6\% of the participants (68) stated that they are mostly comfortable (either very comfortable or comfortable) using it. 45\% of the participants (41) believed that code generated by a device linked to their accounts is more secure than OTPs.
 
The most valuable aspect of 2FA was found to be the added security layer, as reported by 68 (75.6\%) participants, while 52 (57.8\%) participants appreciated the increased control it offered. Twenty-two (24.4\%) participants indicated that 2FA provided peace of mind by not having to worry about account leaks. 
Of the 87 (96.7\%) participants who observed browser prompts for saving passwords, 63 used 2FA in some form. Interestingly, 23 of these participants who used 2FA opted not to use the browser’s built-in password managers.

\section{Discussion} 
\label{discussion}
Next, we discuss our findings from the study concerning Indians' habits and perceptions of authentication tools. 
\changed{\subsection{Password habits}}
Our findings indicate a common security shortfall in user behaviour where convenience (e.g., password reuse) is valued over uniqueness and complexity. Similar results were reported as part of a previous study~\cite{wang_domino} that performed a large-scale empirical analysis of password reuse of 28.8 million users and their 61.5 million passwords, showing that password reuse was widespread. This could be primarily attributed to a common misconception among the users that their accounts do not contain any sensitive or useful information to be hacked. 
Despite giving importance to staying up to date with security trends, the participants lacked  awareness of good security practices.
More than 75\% of the participants reused  passwords and preferred medium-length passwords while using personal data like names and phone numbers in their passwords.

The majority of our participants preferred not to use a password manager/2FA vendor if they were informed about such vendors experiencing data breaches. However, participants rarely updated themselves with new threats, similar to the findings of Lahcen et al.~\cite{lahcen}.
It is, thus, necessary to educate people about the seriousness of these insecure practices and bridge the divide between their beliefs and habits. 

\changed{\subsection{Password managers}}

Password managers are primarily used for convenience, as evident from the popular reasons chosen by participants for using them. This also aligns with the findings of prior works~\cite{reese2019usability,ZHAO201432,fagan2017investigation} that found that ease of use is a significant factor in password manager adoption among users. The users may not be aware that these managers can help create and remember stronger passwords, thereby preventing brute-force-like attacks on their accounts.

Our study highlights a lack of trust in password managers amongst Indian users, which is also not absolute. Most users do not prefer to store their banking credentials in a password manager, in contrast to their behaviour with social media and e-commerce passwords. This may be because they fear the safety of their financial information and believe that the password manager is sharing sensitive password credentials with their servers or elsewhere. Similar sentiments of not using a password manager in the case of sensitive accounts have been echoed through other studies elsewhere~\cite{pearman2019people, sabrina_bank}. 

\changed{\subsection{Two-factor authentication}}
OTPs are heavily used and preferred as a 2FA method, which can be explained by OTPs being the default 2FA method in most Indian financial systems~\cite{rbi2025vision}, making the majority of people familiar with them and associating them with security. Although phishing in the Indian context has been studied previously~\cite{phishing_international}, they focused more on its psychological and social aspects than their knowledge of phishing. Indian users’ awareness of the attacks that can be made on OTPs, like SIM swapping and social engineering, needs to be studied further. This will help us get a better idea of how to mitigate against the attacks and how to increase awareness about it, especially when Indians are more susceptible to phishing~\cite{phishing_international}.

In general, Indian users feel positive about using 2FA due to the added security and comfort, in contrast to password managers, which are less trusted, especially for sensitive applications. 
However, the use of 2FA techniques like OTPs was criticized as they prevented access in areas without a network. An alternative would be to provide multiple ways for 2FA authentication, e.g., providing them OTPs over their mobile network and emails, which many organizations have started adopting. This would allow the users to embrace 2FA further and provide them access to online services even if they do not have access to some 2FA tools. 

Another important issue reported by the participants was that 2FA-based authentication was not particularly user-friendly for people of old-age. It would be interesting to study this further as we had limited data points in this demographic. 
People also felt that 2FA recovery techniques need to be made more user-friendly by providing multiple backup options to avoid getting locked out of their accounts in extreme situations.

\changed{\subsection{Other methods}}
\added{Instead of relying solely on passwords and second-factor authentication, integrating other technologies can enhance the user experience. \adding{Prior research indicates that methods such as risk-based authentication, implicit authentication, and continuous authentication can be effective~\cite{morethengoodpass,rba,impa,ca,PMID:34502865}. Additionally, these methods are also stated to be explored by the Reserve Bank of India~\cite{rbi2025vision}.}
Future studies could explore the adoption and usability of these technologies in conjunction with traditional 2FA among Indian users.}

\section{Conclusion} 
\label{conclusion}
In this work, we investigate the habits of Indian users when authenticating themselves online. We conducted a survey with 90 participants residing in India to understand their habits and perceptions with respect to passwords, password managers and two-factor authentication techniques. Our findings indicate that Indians are aware of the need to follow security practices but do not do so in real life. All the participants reported using password managers and 2FA in some or other form daily, but only if it was the default. Convenience was a major reason for using password managers, while the users were happy to use 2FA as an additional security layer. Training people on using these tools may increase their chances of adopting them although the users did not complain about having some of these as default requirements. The majority of the participants did not trust password managers with credentials for high-sensitivity applications but felt safer when using 2FA for similar online activities.  

\section{Acknowledgments}

We would like to thank the anonymous reviewers for their feedback on our paper. This work was supported in part by the Science and Engineering Research Board (SERB) via grant SRG/2023/000075.

\bibliographystyle{IEEEtran}
\bibliography{sample-base}

\end{document}